# Electromagnetic force and torque in ponderable media

**Masud Mansuripur**

College of Optical Sciences, The University of Arizona, Tucson, Arizona 85721
masud@optics.arizona.edu

**Abstract**: Maxwell's macroscopic equations combined with a generalized form of the Lorentz law of force are a complete and consistent set of equations. Not only are these five equations fully compatible with special relativity, they also conform with conservation laws of energy, momentum, and angular momentum. We demonstrate consistency with the conservation laws by showing that, when a beam of light enters a magnetic dielectric, a fraction of the incident linear (or angular) momentum pours into the medium at a rate determined by the Abraham momentum density, $\boldsymbol{E} \times \boldsymbol{H}/c^2$, and the group velocity $V_g$ of the electromagnetic field. The balance of the incident, reflected, and transmitted momenta is subsequently transferred to the medium as force (or torque) at the leading edge of the beam, which propagates through the medium with velocity $V_g$. Our analysis does not require "hidden" momenta to comply with the conservation laws, nor does it dissolve into ambiguities with regard to the nature of electromagnetic momentum in ponderable media. The linear and angular momenta of the electromagnetic field are clearly associated with the Abraham momentum, and the phase and group refractive indices ($n_p$ and $n_g$) play distinct yet definitive roles in the expressions of force, torque, and momentum densities.

**OCIS codes**: (260.2110) Electromagnetic theory ; (140.7010) Trapping ; (160.3918) Metamaterials.

## 1. Introduction

Standard textbooks on electromagnetism tend to treat the macroscopic equations of Maxwell as somehow inferior to their microscopic counterparts [1,2]. This is due to the fact that, for real materials, polarization and magnetization densities ***P*** and ***M*** are defined as averages over small volumes that must nevertheless contain a large number of atomic dipoles. Consequently, the macroscopic ***E***, ***D***, ***H*** and ***B*** fields are regarded as spatial averages of the "actual" fields; without averaging, these fields would be wildly fluctuating on the scale of atomic dimensions. (The actual fields, of course, are presumed to be well-defined at all points in space and time.) There is also a tendency to elevate ***E*** and ***B*** to the status of "fundamental," while treating ***D*** and ***H*** as secondary or "derived" fields.

This is an unfortunate state of affairs, considering that the macroscopic equations of Maxwell are a complete and self-consistent set, provided that the fields are treated as precisely-defined mathematical entities, i.e., without attempting to associate ***P*** and ***M*** with the properties of real materials. Stated differently, if material media consisted of dense collections of point dipoles, then any volume of the material, no matter how small, would contain an infinite number of such dipoles, eliminating thereby the need for the introduction of macroscopic averages into Maxwell's equations. Also, since in their simplest form, the macroscopic equations contain all four of the ***E***, ***D***, ***H***, ***B*** fields, one should perhaps resist the temptation to designate some of these as more fundamental than others. Tellegen [3] has promoted the idea that ***E***, ***D***, ***H*** and ***B*** should be regarded as equally important physical entities, a point of view with which we agree.

Constitutive relations equate ***P*** with $D - \varepsilon_o E$ and ***M*** with $B - \mu_o H$, thus allowing ***P*** and ***M*** to be designated as secondary fields. Electric and magnetic energy densities and the Poynting vector may now be written as $\mathcal{E}_e = E \cdot D$, $\mathcal{E}_m = H \cdot B$, and $S = E \times H$, respectively, without the need to explain away the appearance of "derived" fields ***D*** and ***H*** in the expressions pertaining to a most fundamental physical entity. [We note in passing that, in deriving Poynting's theorem, the assumed rate of change of energy density

$$\partial \mathcal{E} / \partial t = E \cdot J_{\text{free}} + E \cdot \partial D / \partial t + H \cdot \partial B / \partial t \qquad (1)$$

is, in fact, the *only* postulate of the classical theory concerning electromagnetic energy.]

The fifth fundamental equation of the classical theory, the Lorentz law of force $F = q(E + V \times B)$, expresses the force experienced by a particle of charge $q$ moving with velocity ***V*** through an electromagnetic field. It is fairly straightforward to derive from this law the force and torque exerted on an electric dipole ***p*** (or the force and torque densities exerted on the polarization ***P***). However, the Lorentz law is silent on the question of force/torque experienced by a magnetic dipole ***m*** in the presence of an electromagnetic field. Traditionally, magnetic dipoles have been treated as Amperian current loops, and the force and torque exerted upon them have been derived from the standard Lorentz law by considering the loop's current as arising from circulating electric charges. The problem with this approach is that, when examining the propagation of electromagnetic waves through magnetic media, one finds that linear and angular momenta are *not* conserved. Shockley [4] has famously called attention to the problem of "hidden" momentum within magnetic materials. Fortunately, it is possible to extend the Lorentz law to include the electromagnetic forces on both electric and magnetic dipoles in a way that is consistent with the conservation of energy, momentum, and angular momentum. This extension of the Lorentz law has been attempted a few times during the past forty years, each time from a different perspective but always resulting in essentially the same generalized form of the force law [4-10]. It is now possible to claim that we finally possess a generalized Lorentz law which, in conjunction with Maxwell's macroscopic equations, is fully consistent with the conservation laws of physics.

The goal of the present paper is to demonstrate the consistency of the generalized Lorentz law with conservation of linear and angular momenta. For the most part we will confine our attention to the case of homogeneous, isotropic, linear, and transparent media specified by their relative permittivity $\varepsilon(\omega)$ and permeability $\mu(\omega)$, although a case involving a birefringent medium is discussed in section 5 as well. We derive expressions for the total force and torque exerted on magnetic dielectrics, thus clarifying the reasons behind the traditional division of linear and angular momenta into electromagnetic and mechanical parts. Our methods should be applicable not only to transparent media, whose $\varepsilon(\omega)$ and $\mu(\omega)$ are real-valued, but also to absorbing media, where at least one of these parameters is complex.

We emphasize at the outset that the results of the present analysis, when specialized to non-magnetic and/or non-dispersive media, are in complete agreement with our previous publications as well as with the results of Loudon and his co-workers reported in [11-14]. The present paper describes general methods of calculating electromagnetic force and torque in homogenous, linear, magnetic dielectrics. The field imparts some of its momentum to the host medium, typically at the leading or trailing edges of a light pulse, at the side-walls of a finite-diameter beam, or at the surfaces and interfaces that separate homogeneous media of differing optical constants. These forces and torques convert some fraction of the electromagnetic momentum into mechanical momentum of the host (or vice-versa), while some of the initial momentum continues to exist in electromagnetic form. Under all the circumstances considered, the total linear and angular momenta of the light-matter system, comprising electromagnetic and mechanical contributions, are conserved.

**2. Force and torque exerted on electric and magnetic dipoles by the electromagnetic field**

In a recent publication [10] we derived the following generalized expression for the Lorentz force density in a homogeneous, linear, isotropic medium specified by its $\mu$ and $\varepsilon$ parameters:

$$\boldsymbol{F}_1(\boldsymbol{r},t) = (\boldsymbol{P}\cdot\nabla)\boldsymbol{E} + (\boldsymbol{M}\cdot\nabla)\boldsymbol{H} + (\partial\boldsymbol{P}/\partial t)\times\mu_o\boldsymbol{H} - (\partial\boldsymbol{M}/\partial t)\times\varepsilon_o\boldsymbol{E}. \qquad (2)$$

Similar expressions have been derived by others (see, for example, Hansen and Yaghjian [8]). Our focus, however, has been the generalization of the Lorentz law in a way that is consistent with Maxwell's equations, with the principles of special relativity, and with the conservation laws. In conjunction with Eq. (2), Maxwell's equations in the MKSA system of units are:

$$\nabla\cdot\boldsymbol{D} = \rho_{\text{free}}, \qquad \nabla\times\boldsymbol{H} = \boldsymbol{J}_{\text{free}} + \partial\boldsymbol{D}/\partial t, \qquad \nabla\times\boldsymbol{E} = -\partial\boldsymbol{B}/\partial t, \qquad \nabla\cdot\boldsymbol{B} = 0. \qquad (3\text{a})$$

In these equations, electric displacement $\boldsymbol{D}$ and magnetic induction $\boldsymbol{B}$ are related to the polarization density $\boldsymbol{P}$ and magnetization density $\boldsymbol{M}$ via the constitutive relations

$$\boldsymbol{D} = \varepsilon_o\boldsymbol{E} + \boldsymbol{P} = \varepsilon_o(1+\chi_e)\boldsymbol{E} = \varepsilon_o\varepsilon\boldsymbol{E}, \qquad \boldsymbol{B} = \mu_o\boldsymbol{H} + \boldsymbol{M} = \mu_o(1+\chi_m)\boldsymbol{H} = \mu_o\mu\boldsymbol{H}. \qquad (3\text{b})$$

In what follows, the medium will be assumed to have neither free charges nor free currents (i.e., $\rho_{\text{free}}=0$, $\boldsymbol{J}_{\text{free}}=0$). Our homogeneous, linear, isotropic media will be assumed to be fully specified by their permittivity $\varepsilon = \varepsilon' + i\varepsilon''$ and permeability $\mu = \mu' + i\mu''$. Any loss of energy in such media will be associated with $\varepsilon''$ and $\mu''$, which, by convention, are $\geq 0$. The real parts of $\varepsilon$ and $\mu$, however, may be positive or negative; in particular, in negative-index media, $\varepsilon' < 0$ and $\mu' < 0$. Using simple examples that are amenable to exact analysis, we have shown in a previous publication [10] that Eqs. (2) and (3) lead to a precise balance of momentum when all relevant forces, especially those at the boundaries, are properly taken into account. A major concern of the present paper is the extension of these arguments to prove the conservation of angular momentum.

In [10] and elsewhere, we have considered an alternative formulation of the generalized Lorentz law, where bound electric and magnetic charge densities $\rho_e = -\nabla\cdot\boldsymbol{P}$ and $\rho_m = -\nabla\cdot\boldsymbol{M}$ directly experience the force of the $\boldsymbol{E}$ and $\boldsymbol{H}$ fields. The alternative formula is

$$\boldsymbol{F}_2(\boldsymbol{r},t) = -(\nabla\cdot\boldsymbol{P})\boldsymbol{E} - (\nabla\cdot\boldsymbol{M})\boldsymbol{H} + (\partial\boldsymbol{P}/\partial t)\times\mu_o\boldsymbol{H} - (\partial\boldsymbol{M}/\partial t)\times\varepsilon_o\boldsymbol{E}. \qquad (4)$$

As far as the total force exerted on a given volume of material is concerned, Eqs. (2) and (4) can be shown to yield identical results provided that forces at the boundaries are properly treated in each case in accordance with the corresponding force equation [13, 14]. The force distribution throughout the volume, of course, will depend on which formulation is used, but when integrated over the volume of interest, the two distributions always yield identical values for total force. With regard to torque, the situation is somewhat different. If Eq. (2) is used as the force expression, then torque density will be

$$T_1(r,t) = r \times F_1(r,t) + P(r,t) \times E(r,t) + M(r,t) \times H(r,t). \tag{5}$$

On the other hand, the force expression of Eq. (4) is all that is needed for the calculation of torque, namely,

$$T_2(r,t) = r \times F_2(r,t). \tag{6}$$

Again, Eqs. (5) and (6) can be shown to yield identical results for the total torque on a given volume of material, even though the predicted torque density distributions are usually different in the two formulations. The proof of equivalence of total force (and total torque) for the two formulations was originally given by Barnett and Loudon [13, 14]. Subsequently, we extended their proof to cover the case of objects immersed in a liquid [15]. In our proof, we stated that $P \times E$ (and, by analogy, $M \times H$) will be zero in isotropic media and, therefore, the additional terms in Eq. (5) need not be considered. This statement, while valid in some cases, is generally incorrect. In other words, for the total torques in the two formulations to be identical, Eq. (5) must definitely contain the $P \times E$ and $M \times H$ contributions.

In birefringent media, of course, $P(r,t)$ is not always parallel to $E(r,t)$, nor is $M(r,t)$ parallel to $H(r,t)$, thus making it obvious that the $P \times E$ and $M \times H$ terms in Eq. (5) are indeed necessary. Even in the case of isotropic media, these terms are needed in many circumstances, because $P$ and $E$ (or $M$ and $H$) could assume differing orientations. For instance, in a time-harmonic (i.e., single-frequency) circularly polarized electromagnetic field, $P$ lags behind the rotating $E$-field when $\varepsilon$ is complex; similarly, $M$ lags behind the rotating $H$-field when $\mu$ is complex. In general, therefore, when computing the torque in accordance with Eq. (5), one must beware of the possibility that absorption, dispersion, or birefringence could all create conditions under which $P$ and $E$ (or $M$ and $H$) will have differing orientations. One such situation will be encountered in section 5 below.

## 3. Relation between the Lorentz force and the forces obtained from energy gradients

The energy of a single electric dipole $p$ immersed in an electromagnetic field is $\mathcal{E}_p = p \cdot E$, while that of a single magnetic dipole $m$ is $\mathcal{E}_m = m \cdot H$. The force experienced by these dipoles is thus expected to be [1, 2]:

$$f_p = \nabla(p \cdot E) = (p \cdot \nabla)E + p \times (\nabla \times E) = (p \cdot \nabla)E - p \times \partial B/\partial t, \tag{7a}$$

$$f_m = \nabla(m \cdot H) = (m \cdot \nabla)H + m \times (\nabla \times H) = (m \cdot \nabla)H + m \times \partial D/\partial t. \tag{7b}$$

Note that $p$ and $m$ in these expressions represent individual dipoles and, therefore, the gradient operator acts only on the fields $E$ and $H$ in which the dipoles are immersed. The force in Eq. (2), however, is not exactly the same as that predicted by these energy gradients. One can rewrite Eq. (2) in terms of energy gradients as follows:

$$F_1(r,t) = \nabla(P \cdot E + M \cdot H) + \partial(D \times B - E \times H/c^2)/\partial t. \tag{8}$$

We emphasize once again that the gradient operator in Eq. (8) acts on the $E$ and $H$ fields only; $P$ and $M$ must be treated as locally constant fields. The bottom line is that the effective force experienced by $P$ and $M$ has an extra term given by the time derivative of $D \times B - (E \times H)/c^2$, which happens to be the difference between the Minkowski and Abraham momentum densities [16]. Once the fields settle into a single-frequency (i.e., time-harmonic) oscillation, the contribution to average force of the time-derivative term in Eq. (8) vanishes, leaving the gradient term as the effective force. However, in the presence of transient events, the time-derivative term must be taken into account, otherwise one will end up with a certain amount of hidden momentum in the system.

## 4. Pulse of light entering a transparent, homogeneous, isotropic medium

In [17] we studied the transfer of angular momentum from a circularly polarized plane-wave to a semi-infinite isotropic dielectric using Eq. (6). Here we investigate a more general version of the problem involving a transparent magnetic medium using the alternative formulation of the Lorentz law given in Eqs. (2) and (5). Shown in Fig. 1 is a wide (but finite-diameter) light pulse entering a semi-infinite slab at normal incidence. The leading edge of the pulse propagates along the $z$-axis, exerting a force and a torque on the medium, which account, respectively, for the *mechanical* linear and angular momenta of the light inside the medium. The medium is transparent, isotropic, and dispersionless, so that, at all points along the beam's path, $P$ is parallel to $E$ and $M$ is parallel to $H$. Also, at normal incidence, there will be no forces or torques exerted at the entrance facet (when the beam diameter is sufficiently large). The entire force and torque will thus arise from the action of the leading edge of the light pulse on electric and magnetic dipoles in accordance with Eqs. (2) and (5).

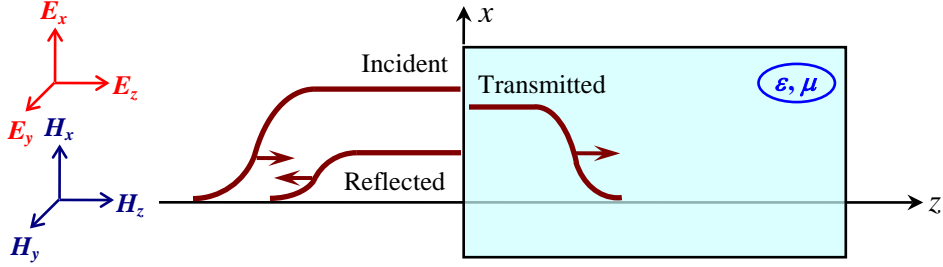

Fig. 1. A semi-infinite slab having material parameters ($\varepsilon, \mu$) is illuminated at normal incidence with an elliptically-polarized, finite diameter light pulse. The incident $E$ and $H$ amplitudes are ($E_{xo}, E_{yo}, E_{zo}$) and ($H_{xo}, H_{yo}, H_{zo}$), respectively. The isotropic material is transparent and free from dispersion, i.e., ($\varepsilon, \mu$) are real-valued and frequency-independent. The refractive index $n = \sqrt{\mu\varepsilon}$ thus determines the speed of the leading edge of the pulse within the material as $c/n$.

We begin by rewriting the force equation, Eq. (2), for a transparent, isotropic, dispersion-less medium in terms of the real-valued field amplitudes $E(r, t)$ and $H(r, t)$. This is possible because $\varepsilon$ and $\mu$ are real-valued and frequency-independent. We set $P(r,t) = \varepsilon_o(\varepsilon - 1)E(r,t)$ and $M(r,t) = \mu_o(\mu - 1)H(r,t)$, use Maxwell's curl equations to replace some of the space-derivatives with time-derivatives, combine various terms, and find the following equivalent of Eq. (2):

$$F_1(r,t) = \tfrac{1}{2}\varepsilon_o(\varepsilon-1)\nabla(E_x^2 + E_y^2 + E_z^2) + \tfrac{1}{2}\mu_o(\mu-1)\nabla(H_x^2 + H_y^2 + H_z^2) + (\mu\varepsilon - 1)\frac{\partial}{\partial t}(E \times H/c^2). \quad (9)$$

Let us first compute the total force exerted by the transmitted portion of the pulse on the semi-infinite slab. Integrating the force density $F_1(r, t)$ over the volume of the slab, we find that the gradient terms along $\hat{x}$ and $\hat{y}$ vanish, leaving only the gradient term along $\hat{z}$. The third term in Eq. (9) contains a time derivative, but since the beam travels with a speed of $c/n$ along the $z$-axis, $\partial/\partial t$ can be replaced with $-(c/n)\partial/\partial z$, where $n = \sqrt{\mu\varepsilon}$ is the refractive index of the material. (Note that lack of dispersion makes the group velocity equal to the phase velocity.) The integrated force is thus given by

$$F_1(t) = -\tfrac{1}{2}\big[\varepsilon_o(\varepsilon-1)\iint_{-\infty}^{\infty}(E_x^2+E_y^2+E_z^2)\big|_{z=0^+}\mathrm{d}x\mathrm{d}y + \mu_o(\mu-1)\iint_{-\infty}^{\infty}(H_x^2+H_y^2+H_z^2)\big|_{z=0^+}\mathrm{d}x\mathrm{d}y\big]\hat{z}$$
$$+ (c/n)(\mu\varepsilon-1)\iint_{-\infty}^{\infty}(E\times H/c^2)\big|_{z=0^+}\mathrm{d}x\mathrm{d}y. \quad (10)$$

Once the leading edge is sufficiently advanced inside the material, the field amplitudes at $z = 0^+$ stabilize and assume sinusoidal behavior. The $z$-component of the force can then be time-averaged over one period of oscillation. The beam diameter is large enough that the Fresnel reflection coefficient at normal incidence,

$$r = (1 - \sqrt{\varepsilon/\mu})/(1 + \sqrt{\varepsilon/\mu}), \quad (11)$$

is all that is needed to determine the reflected and transmitted *E*- and *H*-field amplitudes. At $z = 0^+$, immediately beneath the surface, the amplitudes of $E_x, E_y, B_z$ are $(1+r)$ times the corresponding incident amplitudes, while those of $H_x, H_y, D_z$ are $(1-r)$ times the corresponding incident amplitudes. This is a consequence of the symmetry of reflection as well as the continuity of $E_\parallel$, $H_\parallel$, $B_\perp$ and $D_\perp$ field components. Recognizing that the beam's cross-sectional area is large, we ignore $E_z$ and $H_z$ at $z = 0^+$, then treat the remaining field amplitudes as uniform in the *xy*-plane at $z = 0^+$. Normalizing the integrated force by the cross-sectional area of the beam, the time-averaged force per unit area is found to be

$$\langle F_z \rangle = -\tfrac{1}{4}[\varepsilon_0(\varepsilon-1)(E_x^2+E_y^2)+\mu_0(\mu-1)(H_x^2+H_y^2)-2(cn)^{-1}(\mu\varepsilon-1)(E_xH_y-E_yH_x)]\big|_{(x=0,y=0,z=0+)}$$

$$= -\tfrac{1}{4}\varepsilon_0[(\varepsilon-1)(1+r)^2+(\mu-1)(1-r)^2-2n^{-1}(\mu\varepsilon-1)(1-r^2)](E_{xo}^2+E_{yo}^2)$$

$$= \tfrac{1}{4}\varepsilon_0 n^{-1}(\mu+\varepsilon-2)(1-r^2)(E_{xo}^2+E_{yo}^2). \qquad (12)$$

This time-averaged force is exerted on the slab by the leading edge of the beam as it propagates within the medium. In addition, electromagnetic (i.e., Abraham) momentum pours into the slab with a volume density of $\boldsymbol{p}_{EM}=\boldsymbol{E}\times\boldsymbol{H}/c^2$. Since the leading edge moves a distance of *c*/*n* in one second, the electromagnetic momentum (per unit area per second) delivered to the medium is $(cn)^{-1}\langle\boldsymbol{E}\times\boldsymbol{H}\rangle = \tfrac{1}{2}\varepsilon_0 n^{-1}(1-r^2)(E_{xo}^2+E_{yo}^2)\hat{\boldsymbol{z}}$. Adding this to $\langle F_z \rangle$ of Eq. (12) yields the total rate of flow of linear momentum into the slab as $\tfrac{1}{2}\varepsilon_0(1+r^2)(E_{xo}^2+E_{yo}^2)$. The final result is exactly equal to the rate of flow of incident plus reflected momenta in the free space, thus establishing the conservation of linear momentum.

Extending the analysis to dispersive media (see the appendices) reveals the same general behavior, except that the leading edge of the beam now propagates with the group velocity $V_g = c/n_g$. The electromagnetic momentum density will still be given by the Abraham formula, $\boldsymbol{p}_{EM}=\boldsymbol{E}\times\boldsymbol{H}/c^2$, and the rate of flow of energy into the medium will still be determined by the Poynting vector $\boldsymbol{S} = \boldsymbol{E}\times\boldsymbol{H}$. If we then proceed to assume that the energy contained in a given volume of the material is $N\hbar\omega$, with $\hbar$ being the reduced Planck constant, $\omega$ the angular frequency of the light, and *N* the number of photons, the Abraham momentum per photon turns out to be $\hbar\omega/(n_g c)$. This is the general formula for the photon's *electromagnetic* momentum in a transparent medium having group refractive index $n_g$. (This result applies to all transparent media, irrespective of the phase refractive index $n_p$ being positive or negative.)

Next, we examine the torque exerted by $\boldsymbol{F}_1(\boldsymbol{r},t)$ of Eq. (9) on the semi-infinite slab of Fig. 1. (In transparent dispersionless media $\boldsymbol{P}\|\boldsymbol{E}$ and $\boldsymbol{M}\|\boldsymbol{H}$; therefore, $\boldsymbol{P}\times\boldsymbol{E}$ and $\boldsymbol{M}\times\boldsymbol{H}$ of Eq. (5) do not contribute to the torque.) The total torque is obtained by integrating $\boldsymbol{r}\times\boldsymbol{F}_1(\boldsymbol{r},t)$ over the volume of the material, as follows:

$$\boldsymbol{T}(t) = \tfrac{1}{2}\varepsilon_0(\varepsilon-1)\iiint \boldsymbol{r}\times\nabla(E_x^2+E_y^2+E_z^2)\,dxdydz + \tfrac{1}{2}\mu_0(\mu-1)\iiint \boldsymbol{r}\times\nabla(H_x^2+H_y^2+H_z^2)\,dxdydz$$

$$- (c/n)(\mu\varepsilon-1)\iiint \boldsymbol{r}\times\frac{\partial}{\partial z}(\boldsymbol{E}\times\boldsymbol{H}/c^2)\,dxdydz$$

$$= \tfrac{1}{2}\varepsilon_0(\varepsilon-1)\iiint [(y\tfrac{\partial}{\partial z}-z\tfrac{\partial}{\partial y})\hat{\boldsymbol{x}}+(z\tfrac{\partial}{\partial x}-x\tfrac{\partial}{\partial z})\hat{\boldsymbol{y}}+(x\tfrac{\partial}{\partial y}-y\tfrac{\partial}{\partial x})\hat{\boldsymbol{z}}](E_x^2+E_y^2+E_z^2)\,dxdydz$$

$$+ \tfrac{1}{2}\mu_0(\mu-1)\iiint [(y\tfrac{\partial}{\partial z}-z\tfrac{\partial}{\partial y})\hat{\boldsymbol{x}}+(z\tfrac{\partial}{\partial x}-x\tfrac{\partial}{\partial z})\hat{\boldsymbol{y}}+(x\tfrac{\partial}{\partial y}-y\tfrac{\partial}{\partial x})\hat{\boldsymbol{z}}](H_x^2+H_y^2+H_z^2)\,dxdydz$$

$$- (cn)^{-1}(\mu\varepsilon-1)\iiint \Big\{[y\tfrac{\partial}{\partial z}(E_xH_y-E_yH_x)-z\tfrac{\partial}{\partial z}(E_zH_x-E_xH_z)]\hat{\boldsymbol{x}}$$

$$+ [z\tfrac{\partial}{\partial z}(E_yH_z-E_zH_y)-x\tfrac{\partial}{\partial z}(E_xH_y-E_yH_x)]\hat{\boldsymbol{y}}$$

$$+ [x\tfrac{\partial}{\partial z}(E_zH_x-E_xH_z)-y\tfrac{\partial}{\partial z}(E_yH_z-E_zH_y)]\hat{\boldsymbol{z}}\Big\}\,dxdydz. \qquad (13)$$

Some of the integrals in Eq. (13) vanish because integration of the $\partial/\partial x$ or $\partial/\partial y$ terms takes the integrand beyond the beam's finite diameter, where $E$- and $H$-field intensities are zero. Other integrals end up being zero because of the symmetry of the incident beam around the $z$-axis. (We are assuming circular symmetry around $z$, because the emphasis of the present analysis is on "spin" angular momentum, which arises from the polarization state of the beam, as opposed to "orbital" angular momentum, which is rooted in the circulation of the phase profile around the $z$-axis.) The only part of Eq. (13) that survives integration is the very last line, whose $\partial/\partial z$ terms, when integrated over $z$, yield an expression in terms of the field components at $z = 0^+$. These components may then be time-averaged to yield

$$\begin{aligned}
<T_z> &= \tfrac{1}{2}(cn)^{-1}(\mu\varepsilon - 1)\iint \left[ x(E_z H_x - E_x H_z)\big|_{z=0^+} - y(E_y H_z - E_z H_y)\big|_{z=0^+} \right] dxdy \\
&= \tfrac{1}{2}(cn)^{-1}(\mu\varepsilon - 1)\iint \big\{ x[\varepsilon^{-1}(1-r)^2 E_{zo} H_{xo} - \mu^{-1}(1+r)^2 E_{xo} H_{zo}] \\
&\qquad\qquad\qquad\qquad\qquad - y[\mu^{-1}(1+r)^2 E_{yo} H_{zo} - \varepsilon^{-1}(1-r)^2 E_{zo} H_{yo}] \big\} dxdy \\
&= \tfrac{1}{2}c^{-1}[(\mu\varepsilon - 1)/\mu\varepsilon](1-r^2)\iint [x(E_{zo}H_{xo} - E_{xo}H_{zo}) - y(E_{yo}H_{zo} - E_{zo}H_{yo})] dxdy \\
&= (1-n^{-2})(1-r^2)\iint_{-\infty}^{\infty} (<xS_{yo} - yS_{xo}>/c)\, dxdy. \qquad (14)
\end{aligned}$$

Here $<\boldsymbol{S}_o> = \tfrac{1}{2}\boldsymbol{E}_o \times \boldsymbol{H}_o$ is the time-averaged Poynting vector associated with the incident beam. The integral in the final line of Eq. (14) represents the total incident angular momentum per unit time. The integral is multiplied by $(1-r^2)$, which has the effect of subtracting the angular momentum carried away by the reflected beam. (Note that, unlike linear momentum whose contribution due to reflection adds to the exerted force, angular momentum reverses sign upon reflection and, therefore, its contribution to exerted torque must be subtracted.) The torque $<T_z>$ exerted on the medium by the leading edge of the pulse is thus equal to the net angular momentum influx multiplied by $(1-n^{-2})$. This implies that the electromagnetic (i.e., Abraham) angular momentum flux into the medium is $1/n^2$ times the flux of "incident minus reflected" angular momenta. In other words, to conserve the total angular momentum of $\hbar$ per photon, each photon's *electromagnetic* angular momentum inside the slab must reduce to $\hbar/n^2$.

When the analysis is extended to dispersive media (see the appendices), the aforementioned reduction factor $n^2$ becomes $n_p n_g$, the product of phase and group refractive indices. Thus, upon entering a homogeneous, linear, isotropic, and transparent medium, the electromagnetic angular momentum per photon shrinks by a factor of $n_p n_g$, while, according to Eq. (14), the balance of the incident, reflected and transmitted angular momenta is transferred to the medium as a torque (via the leading edge of the beam). This result applies to negative-index media as well, where the sign of the electromagnetic angular momentum is reversed relative to that of the incident beam (because $n_p < 0$).

## 5. Torque on a birefringent slab

A birefringent slab having material parameters $(\varepsilon_x, \mu_x)$ along the $x$-axis and $(\varepsilon_y, \mu_y)$ along the $y$-axis is depicted in Fig. 2. The semi-infinite slab is illuminated at normal incidence with an elliptically-polarized plane-wave. The incident field amplitudes are $(E_{xo}, E_{yo})$ and $(H_{xo}, H_{yo}) = Z_o^{-1}(E_{xo}, E_{yo})$. The incident beam may be written as the sum of left- and right-circularly-polarized plane-waves (LCP and RCP) as follows:

$$E_{xo}\hat{\boldsymbol{x}} + E_{yo}\hat{\boldsymbol{y}} = \tfrac{1}{2}[(E_{xo} - iE_{yo})\hat{\boldsymbol{x}} + i(E_{xo} - iE_{yo})\hat{\boldsymbol{y}}] + \tfrac{1}{2}[(E_{xo} + iE_{yo})\hat{\boldsymbol{x}} - i(E_{xo} + iE_{yo})\hat{\boldsymbol{y}}]. \qquad (15)$$

The rate of flow of angular momentum (per unit area per unit time) carried by the above beams is $L_{z1} = \tfrac{1}{4}(\varepsilon_o/k_o)|E_{xo} - iE_{yo}|^2$ and $L_{z2} = -\tfrac{1}{4}(\varepsilon_o/k_o)|E_{xo} + iE_{yo}|^2$, respectively. Therefore, the total angular momentum influx is $L_z = L_{z1} + L_{z2} = (\varepsilon_o/k_o)\operatorname{Im}(E_{xo}^* E_{yo})$. The reflected beam has $E$-field amplitudes $r_1 E_{xo}$ and $r_2 E_{yo}$, where

$$r_1 = \frac{\sqrt{\mu_y} - \sqrt{\varepsilon_x}}{\sqrt{\mu_y} + \sqrt{\varepsilon_x}}, \qquad r_2 = \frac{\sqrt{\mu_x} - \sqrt{\varepsilon_y}}{\sqrt{\mu_x} + \sqrt{\varepsilon_y}}. \tag{16}$$

The reflected angular momentum flux being $L_z = (\varepsilon_o/k_o)\,\text{Im}(r_1^* E_{xo}^* r_2 E_{yo})$, the angular momentum per unit area per unit time delivered to the semi-infinite slab will be

$$L_z = (\varepsilon_o/k_o)\,\text{Im}[(1 - r_1^* r_2) E_{xo}^* E_{yo}] = 2(\varepsilon_o/k_o)\,\text{Im}\left\{\frac{[\sqrt{\mu_x \varepsilon_x^*} + \sqrt{\mu_y^* \varepsilon_y}] E_{xo}^* E_{yo}}{(\sqrt{\mu_x} + \sqrt{\varepsilon_y})(\sqrt{\mu_y^*} + \sqrt{\varepsilon_x^*})}\right\}. \tag{17}$$

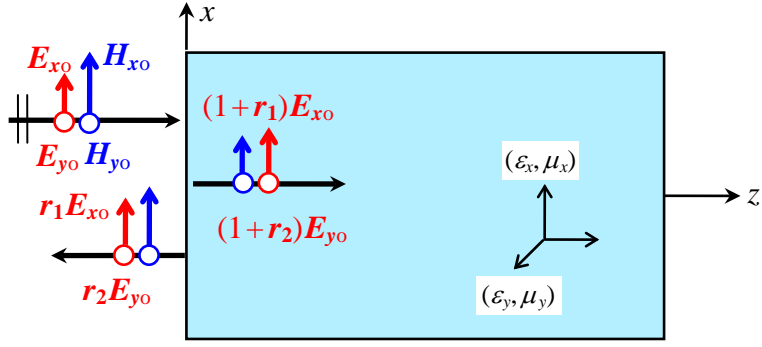

Fig. 2. A semi-infinite birefringent slab having material parameters $(\varepsilon_x, \mu_x)$ along $x$ and $(\varepsilon_y, \mu_y)$ along $y$ is illuminated at normal incidence with an elliptically-polarized plane-wave. The incident $E$ and $H$ amplitudes are related via $(E_{xo}, E_{yo}) = Z_o(H_{yo}, H_{xo})$. The plane-wave defined by $(E_{xo}, H_{yo})$ sees the refractive index $n_1 = \sqrt{\mu_y \varepsilon_x}$ and the reflection coefficient $r_1$; the corresponding parameters for the plane-wave defined by $(E_{yo}, H_{xo})$ are $n_2 = \sqrt{\mu_x \varepsilon_y}$ and $r_2$.

Next we compute directly the torque on the semi-infinite medium. The field amplitudes within the medium are

$$E_x(z, t) = (1 + r_1) E_{xo} \exp(ik_o \sqrt{\mu_y \varepsilon_x}\, z - i\omega t), \tag{18a}$$

$$H_y(z, t) = (1 - r_1) Z_o^{-1} E_{xo} \exp(ik_o \sqrt{\mu_y \varepsilon_x}\, z - i\omega t), \tag{18b}$$

$$E_y(z, t) = (1 + r_2) E_{yo} \exp(ik_o \sqrt{\mu_x \varepsilon_y}\, z - i\omega t), \tag{18c}$$

$$H_x(z, t) = -(1 - r_2) Z_o^{-1} E_{yo} \exp(ik_o \sqrt{\mu_x \varepsilon_y}\, z - i\omega t). \tag{18d}$$

In this problem, the torque arises from the $\boldsymbol{P} \times \boldsymbol{E} + \boldsymbol{M} \times \boldsymbol{H}$ part of Eq.(5); the other contributions, embodied in the $\boldsymbol{r} \times \boldsymbol{F}_1(\boldsymbol{r}, t)$ term, vanish when the latter is expressed in the form of Eq.(13) and its various integrals evaluated, then time-averaged. (Unlike the problem studied in section 4, here we are dealing with the steady-state situation where the leading edge of the beam has already passed through the medium and an exponentially-decaying field along the $z$-axis has been established.) The time-averaged torque per unit surface area experienced by the semi-infinite slab of Fig. 2 is thus given by

$$\langle T_z \rangle = \tfrac{1}{2} \int_0^\infty \text{Re}\,[(\boldsymbol{P}_x \times \boldsymbol{E}_y^* + \boldsymbol{P}_y \times \boldsymbol{E}_x^* + \boldsymbol{M}_x \times \boldsymbol{H}_y^* + \boldsymbol{M}_y \times \boldsymbol{H}_x^*)]\,dz$$

$$= \tfrac{1}{2} \int_0^\infty \text{Re}\,\{\varepsilon_o[(\varepsilon_x^* - 1) - (\varepsilon_y - 1)] E_x^* E_y + \mu_o[(\mu_x - 1) - (\mu_y^* - 1)] H_x H_y^*\}\,dz$$

$$= \tfrac{1}{2} \varepsilon_o \text{Re}\,\{[(\varepsilon_x^* - \varepsilon_y)(1 + r_1^*)(1 + r_2) - (\mu_x - \mu_y^*)(1 - r_1^*)(1 - r_2)]$$
$$\times E_{xo}^* E_{yo} \int_0^\infty \exp[ik_o(\sqrt{\mu_x \varepsilon_y} - \sqrt{\mu_y^* \varepsilon_x^*})\, z]\,dz\}$$

$$= 2(\varepsilon_o/k_o)\,\text{Re}\left\{\frac{i[(\varepsilon_x^* - \varepsilon_y)\sqrt{\mu_x \mu_y^*} - (\mu_x - \mu_y^*)\sqrt{\varepsilon_x^* \varepsilon_y}]}{(\sqrt{\mu_x \varepsilon_y} - \sqrt{\mu_y^* \varepsilon_x^*})(\sqrt{\mu_x} + \sqrt{\varepsilon_y})(\sqrt{\mu_y^*} + \sqrt{\varepsilon_x^*})} E_{xo}^* E_{yo}\right\}$$

$$= 2(\varepsilon_o/k_o)\text{Im}\left\{\frac{\sqrt{\mu_x\varepsilon_x^*}+\sqrt{\mu_y^*\varepsilon_y}}{(\sqrt{\mu_x}+\sqrt{\varepsilon_y})(\sqrt{\mu_y^*}+\sqrt{\varepsilon_x^*})}E_{xo}^*E_{yo}\right\}. \tag{19}$$

An implicit assumption in deriving Eq.(19) is that at least one of the four parameters $\mu_x$, $\mu_y$, $\varepsilon_x$, $\varepsilon_y$ is complex, so that absorption of light within the semi-infinite slab would cause the exponential function in the integrand to approach zero when $z\to\infty$. However, since the term $(\sqrt{\mu_x\varepsilon_y}-\sqrt{\mu_y^*\varepsilon_x^*})$ appearing in the denominator is eventually cancelled out by an identical term in the numerator, the end result is independent of any material absorption.

The final result of Eq.(19) is in complete accord with Eq.(17), that is, the time-averaged torque exerted on the semi-infinite slab is exactly equal to the angular momentum influx carried by the incident beam, minus the angular momentum carried away by the beam reflected at the surface.

## 6. Concluding remarks

The generalized Lorentz law in conjunction with macroscopic Maxwell's equations, Eqs.(3), provide a complete and consistent set of equations that are fully compatible with the laws of conservation of energy and momentum. We have argued that the electromagnetic fields *E*, *D*, *H*, *B* should be treated as fundamental, while polarization and magnetization densities *P* and *M* can be considered as secondary fields derived from the constitutive relations, Eqs.(3b). Two formulations of the generalized Lorentz law are given in Eq.(2) and Eq.(4). These formulas are equally acceptable in the sense that they are both consistent with the conservation laws; moreover, they predict precisely the same *total* force on a given body of material, although the predicted force *distributions* at the surfaces and throughout the volume of the material could be drastically different in the two formulations. If Eq.(2) is used as the generalized law of force, then torque density will be given by Eq.(5). On the other hand, if force is given by Eq.(4), then the torque formula will be Eq.(6).

Two other postulates that need explicit enunciation in the classical theory of electromagnetism are related to energy and momentum. The first postulate declares that the time-rate-of-change of energy density is given by Eq.(1). The second postulate establishes the (linear) momentum density of the electromagnetic field as $\boldsymbol{p}_{EM}=\boldsymbol{E}\times\boldsymbol{H}/c^2$. Once these postulates are accepted, one can easily show that the rate of flow of energy (per unit area per unit time) is given by the Poynting vector $\boldsymbol{S}=\boldsymbol{E}\times\boldsymbol{H}$, that the linear and angular momenta enter and exit a given medium with the group velocity $V_g$, and that the balance of the incident, reflected and transmitted momenta is exerted by the electromagnetic field on the medium as force and torque. These forces and torques are typically concentrated at the edges of the beam and at the surfaces and interfaces of the material medium. In the examples presented in this paper, where the normally-incident beam is fairly uniform, circularly polarized, and has a large diameter, these forces and torques are localized at the leading edge of the beam. [Depending on which form of the Lorentz law is used, i.e., Eq. (2) or Eq. (4), the forces and torques may appear in one part of the beam or another (e.g., at the leading edge of the beam or at the entrance facet of the medium), the force may be compressive at the side-walls of the beam in one formulation and expansive in the other, but, in all cases, the total force (and total torque) exerted on the material body will be found to be exactly the same, no matter which formulation is used.]

Going slightly beyond the classical theory, if one assumes that the electromagnetic energy contained in a given volume of space (or material body) is divided into individual packets of $\hbar\omega$, then the electromagnetic (i.e., Abraham) momentum corresponding to each such bundle of energy (i.e., photon) will be $\hbar\omega/(n_g c)$, where $n_g$ is the group refractive index of the medium. For circularly-polarized light, the spin angular momentum of each photon is $\hbar/(n_p n_g)$, where $n_p$ is the phase refractive index, in agreement with the predictions of Ref. [12]. The balance of linear and angular momenta among the incident, reflected, and transmitted beams is always transferred to the medium in the form of force and torque, which are sometimes identified as "mechanical" momenta of the light beam. When a beam of light enters a material medium from the free space, fractions of its linear and angular momenta remain electromagnetic, while the rest are exerted on the medium in the form of mechanical force and torque. Similarly, when a beam of light emerges from a host medium into the free space, its electromagnetic momenta are augmented by additional momenta that are taken away from the medium, resulting in mechanical backlash (i.e., oppositely directed force and torque) on the medium.

We close by pointing out that our findings do not necessarily contradict the experimental results of [18], where the angular momentum transferred to an antenna at the center of a microwave cavity was found to be the same whether the antenna was placed in the air or immersed in a liquid dielectric. The photon inside the liquid has its electromagnetic angular momentum of $\hbar/(n_p n_g)$, but it is also "dressed" with a certain amount of mechanical angular momentum. What is transferred to the antenna is, in general, a combination of the photon's electromagnetic and mechanical momenta; moreover, the backlash (i.e., momentum transfer to the liquid upon absorption of the photon by the antenna) must also be taken into account.

**Appendix A: Electromagnetic momentum in dispersive media**

Consider the superposition of two finite-diameter beams which are identical in every respect except for a small difference in their temporal frequencies, $\omega_1$ and $\omega_2$. The electric and magnetic field amplitudes of this superposition are expressed in terms of a plane-wave spectrum of spatial frequencies $(k_x, k_y)$ as follows:

$$\boldsymbol{E}(\boldsymbol{r}, t) = \tfrac{1}{2} \sum_{\pm\omega_{1,2}} \iint \boldsymbol{\mathcal{E}}(k_x, k_y, \omega) \exp[\mathrm{i}(k_x x + k_y y + k_z z - \omega t)] \mathrm{d}k_x \mathrm{d}k_y, \tag{A1a}$$

$$\boldsymbol{H}(\boldsymbol{r}, t) = \tfrac{1}{2} \sum_{\pm\omega_{1,2}} \iint \boldsymbol{\mathcal{H}}(k_x, k_y, \omega) \exp[\mathrm{i}(k_x x + k_y y + k_z z - \omega t)] \mathrm{d}k_x \mathrm{d}k_y. \tag{A1b}$$

Here the sum is over $\pm\omega_1$ and $\pm\omega_2$, where $\omega_1$ and $\omega_2$ are two distinct but closely-spaced frequencies. In general, $k_z = (\omega/c)\sqrt{\mu\varepsilon - (ck_x/\omega)^2 - (ck_y/\omega)^2}$. However, since in the present application we are interested in the limit when $k_x$ and $k_y$ are confined to a small region in the vicinity of the origin of the $k_x k_y$-plane, we can safely set $k_z \approx (\omega/c)\sqrt{\mu\varepsilon}$. For the fields to be real-valued it is necessary and sufficient that their Fourier transforms be Hermitian, namely,

$$\boldsymbol{\mathcal{E}}(k_x, k_y, \omega) = \boldsymbol{\mathcal{E}}^*(-k_x, -k_y, -\omega), \qquad \boldsymbol{\mathcal{H}}(k_x, k_y, \omega) = \boldsymbol{\mathcal{H}}^*(-k_x, -k_y, -\omega). \tag{A2}$$

Moreover, if the beam's cross-section in the $xy$-plane is required to be symmetric, say, with respect to the origin, that is, if it is demanded that the field amplitudes remain intact upon switching $(x, y)$ to $(-x, -y)$, we must have

$$\boldsymbol{\mathcal{E}}(k_x, k_y, \omega) = \boldsymbol{\mathcal{E}}(-k_x, -k_y, \omega), \qquad \boldsymbol{\mathcal{H}}(k_x, k_y, \omega) = \boldsymbol{\mathcal{H}}(-k_x, -k_y, \omega). \tag{A3}$$

To ensure the two-frequency superposition of Eqs. (A1) yields a well-defined beat waveform we require the two amplitude-profiles to have equal magnitudes and opposite signs, that is,

$$\boldsymbol{\mathcal{E}}(k_x, k_y, \pm\omega_1) = -\boldsymbol{\mathcal{E}}(k_x, k_y, \pm\omega_2), \qquad \boldsymbol{\mathcal{H}}(k_x, k_y, \pm\omega_1) = -\boldsymbol{\mathcal{H}}(k_x, k_y, \pm\omega_2). \tag{A4}$$

Under these circumstances, the beat-waveform's nodes at $t = 0$ will be located at integer-multiples of $\Delta z \approx 2\pi c/(\omega_2\sqrt{\mu_2 \varepsilon_2} - \omega_1\sqrt{\mu_1 \varepsilon_1})$, and the envelope's travel time between adjacent nodes will be $T = 2\pi/(\omega_2 - \omega_1)$. The beat's group velocity is thus

$$V_g = \frac{c(\omega_2 - \omega_1)}{\omega_2\sqrt{\mu_2 \varepsilon_2} - \omega_1\sqrt{\mu_1 \varepsilon_1}}. \tag{A5}$$

For the beam defined by the above equations, the Poynting vector may be written as follows:

$$S(r, t) = E(r, t) \times H(r, t) = \tfrac{1}{4} \sum \iiiint \mathcal{E}(k_x, k_y, \omega) \times \mathcal{H}(k_x', k_y', \omega') \exp[i(k_x+k_x')x]$$
$$\times \exp[i(k_y+k_y')y] \exp[i(k_z+k_z')z] \exp[-i(\omega+\omega')t] \, dk_x dk_y dk_x' dk_y'. \tag{A6}$$

Integrating $S(r,t)$ over the beam's cross-sectional area $A$ in the $xy$-plane, normalizing by $A$, and using the identity

$$\int_{-\infty}^{\infty} \exp[i(k+k')\zeta] \, d\zeta = \delta(k+k'), \tag{A7}$$

where $\delta(k)$ is Dirac's delta function, the beam's Abraham momentum density turns out to be

$$p_{EM}(z=0, t) = (1/Ac^2) \iint_{-\infty}^{\infty} S(x, y, z=0, t) \, dx \, dy$$
$$= (4Ac^2)^{-1} \sum \iint_{-\infty}^{\infty} \mathcal{E}(k_x, k_y, \omega) \times \mathcal{H}(-k_x, -k_y, \omega') \exp[-i(\omega+\omega')t] \, dk_x dk_y. \tag{A8}$$

Notice that $\omega$ and $\omega'$ can each assume four different values, namely, $\pm\omega_1$ and $\pm\omega_2$, for a total of 16 terms in the above summation. When time-averaging is performed over Eq. (A8), the only terms for which $(1/T)\int_0^T \exp[-i(\omega+\omega')t] \, dt \neq 0$ will be those with $\omega + \omega' = 0$, namely, $(\omega, \omega') = (-\omega_1, \omega_1), (-\omega_2, \omega_2), (\omega_1, -\omega_1),$ and $(\omega_2, -\omega_2)$. Therefore,

$$\langle p_{EM}(z=0, t) \rangle = (2Ac^2)^{-1} \sum_{\omega_{1,2}} \text{Real} \left\{ \iint_{-\infty}^{\infty} \mathcal{E}(k_x, k_y, \omega) \times \mathcal{H}^*(k_x, k_y, \omega) \, dk_x dk_y \right\}. \tag{A9}$$

**Appendix B: Electromagnetic angular momentum in dispersive media**

For the beam defined by Eqs. (A1), the volume density of the $z$-component of angular momentum, $L_z(z=0, t)$, is given by

$$L_z(z=0, t) = (Ac^2)^{-1} \iint_{-\infty}^{\infty} [x S_y(x, y, z=0, t) - y S_x(x, y, z=0, t)] \, dx \, dy. \tag{B1}$$

Using the identity

$$\int_{-\infty}^{\infty} \zeta \exp[i(k+k')\zeta] \, d\zeta = -i\delta'(k+k'), \tag{B2}$$

where $\delta'(k)$ is the derivative of $\delta(k)$ with respect to $k$, we find, upon time-averaging,

$$\langle L_z(z=0, t) \rangle = (4iAc^2)^{-1} \sum_{\pm\omega_{1,2}} \iint [\mathcal{H}_x(k_x, k_y, \omega)(\partial/\partial k_x)\mathcal{E}_z^*(k_x, k_y, \omega) - \mathcal{H}_z(k_x, k_y, \omega)(\partial/\partial k_x)\mathcal{E}_x^*(k_x, k_y, \omega)$$
$$- \mathcal{H}_z(k_x, k_y, \omega)(\partial/\partial k_y)\mathcal{E}_y^*(k_x, k_y, \omega) + \mathcal{H}_y(k_x, k_y, \omega)(\partial/\partial k_y)\mathcal{E}_z^*(k_x, k_y, \omega)] \, dk_x dk_y. \tag{B3}$$

We now replace the $\mathcal{H}$-field components of Eq. (B3) with their equivalents in terms of $\mathcal{E}_x, \mathcal{E}_y, \mathcal{E}_z$ using Maxwell's 3rd equation, namely,

$$k_y \mathcal{E}_z - k_z \mathcal{E}_y = \omega\mu_o\mu\mathcal{H}_x, \quad k_z \mathcal{E}_x - k_x \mathcal{E}_z = \omega\mu_o\mu\mathcal{H}_y, \quad k_x \mathcal{E}_y - k_y \mathcal{E}_x = \omega\mu_o\mu\mathcal{H}_z. \tag{B4}$$

We then use Maxwell's 1st equation, $k_x \mathcal{E}_x + k_y \mathcal{E}_y + k_z \mathcal{E}_z = 0$, to substitute for $\mathcal{E}_z$ in terms of $\mathcal{E}_x$ and $\mathcal{E}_y$. [Note that $k_z = (\omega/c)\sqrt{\mu\varepsilon - (ck_x/\omega)^2 - (ck_y/\omega)^2}$ is a function of $k_x$ and $k_y$; therefore, $\partial k_z/\partial k_x$ and $\partial k_z/\partial k_y$ must be included when evaluating Eq. (B3).] Upon algebraic manipulations we find

$$\langle L_z(z=0, t) \rangle = (\varepsilon_o/2A) \sum_{\omega_{1,2}} \iint_{-\infty}^{\infty} (\omega\mu k_z^2)^{-1} \text{Imag} \Big\{ 2(\omega/c)^2 \mu\varepsilon \, \mathcal{E}_x^* \mathcal{E}_y + k_y[(\omega/c)^2\mu\varepsilon - k_y^2] \mathcal{E}_x \partial\mathcal{E}_x^*/\partial k_x$$
$$+ k_y[(\omega/c)^2\mu\varepsilon - k_x^2] \mathcal{E}_y \partial\mathcal{E}_y^*/\partial k_x + k_x k_y^2 \mathcal{E}_x \partial\mathcal{E}_y^*/\partial k_x + k_x k_y^2 \mathcal{E}_y \partial\mathcal{E}_x^*/\partial k_x - k_x[(\omega/c)^2\mu\varepsilon - k_x^2] \mathcal{E}_y \partial\mathcal{E}_y^*/\partial k_y$$
$$- k_x[(\omega/c)^2\mu\varepsilon - k_y^2] \mathcal{E}_x \partial\mathcal{E}_x^*/\partial k_y - k_x^2 k_y \mathcal{E}_y \partial\mathcal{E}_x^*/\partial k_y - k_x^2 k_y \mathcal{E}_x \partial\mathcal{E}_y^*/\partial k_y \Big\} \, dk_x dk_y. \tag{B5}$$

Equation (B5) is an exact expression for the time-averaged angular momentum density (including contributions from both spin and orbital) in a transparent medium specified by $\varepsilon(\omega)$ and $\mu(\omega)$. Presently we are interested only in spin angular momentum, so we confine our attention to the case of circularly-polarized light where $\mathcal{E}_y(k_x,k_y,\omega_{1,2}) = \mathrm{i}\mathcal{E}_x(k_x,k_y,\omega_{1,2})$, with $\mathcal{E}_x$ being a real-valued function of $(k_x,k_y)$ for both $\omega_1$ and $\omega_2$. Under these circumstances, those terms of Eq.(B5) that contain derivatives with respect to $k_x$ or $k_y$ either vanish (because real-valued functions have no imaginary parts) or cancel each other out. Moreover, $k_z \approx (\omega/c)\sqrt{\mu\varepsilon}$, assuming $E_x$ and $E_y$ are smooth functions of $(x,y)$, and that the beam's cross-sectional area $A$ is large compared to a wavelength. Equation (B5) thus simplifies as follows:

$$<L_z(z=0,t)> \approx (\varepsilon_o/A)\sum_{\omega_{1,2}}(\omega\mu)^{-1}\iint_{-\infty}^{\infty}|\mathcal{E}_x(k_x,k_y,\omega)|^2\,dk_x dk_y. \tag{B6}$$

The two frequencies are seen to contribute to $<L_z>$ independently of each other. According to Parseval's theorem, the integrated $|\mathcal{E}_x|^2$ in the $k_x k_y$-plane is equal to the integrated $|E_x|^2$ over the beam's cross-sectional area in the $xy$-plane. The light beam's energy density in a dispersive medium is $<S_z>/V_g = Z_o^{-1}\sqrt{\varepsilon/\mu}|E_x|^2/V_g$, where $V_g = c/n_g$ is the beam's group velocity. The spin angular momentum in a given volume is thus equal to the energy content of the volume divided by $n_g n_p \omega$, where $n_p = \sqrt{\mu\varepsilon}$ is the phase refractive index. (In a negative-index material, $n_p$ is negative and, therefore, the spin angular momentum must change sign upon entering from the free space.)

Suppose a circularly-polarized plane-wave having amplitude $E_{xo}(\hat{\boldsymbol{x}}+\mathrm{i}\hat{\boldsymbol{y}})$ arrives at normal incidence at the surface of a transparent medium specified by $\varepsilon(\omega)$ and $\mu(\omega)$. Using Fresnel's reflection coefficient given by Eq.(11), we find the $E$-field amplitude immediately inside the medium to be $E_x = (1+r)E_{xo} = 2E_{xo}/(1+\sqrt{\varepsilon/\mu})$. The spin angular momentum density in the medium will then be equal to the incident angular momentum density times $(1-r^2)/\sqrt{\mu\varepsilon}$. The factor $(1-r^2)$, of course, accounts for the loss of angular momentum upon reflection at the surface. Whereas in the incidence space the angular momentum flows at the vacuum speed of light $c$, inside the medium the propagation speed is $c/n_g$, causing the spin angular momentum contained in a given length of the beam to drop by a factor of $n_g\sqrt{\mu\varepsilon}$ upon entering the medium. As will be shown in Appendix D below, the difference between the incident and transmitted angular momenta is imparted to the medium as torque.

**Appendix C: Force exerted on a dispersive slab**

Starting with the generalized Lorentz law of Eq.(2), the force exerted by the transmitted beam on the semi-infinite slab of Fig. 1 can be calculated as follows:

$$F_z(x,y,z,t) = P_x(\partial E_z/\partial x) + P_y(\partial E_z/\partial y) + P_z(\partial E_z/\partial z) + (\partial P_x/\partial t)\mu_o H_y - (\partial P_y/\partial t)\mu_o H_x$$
$$+ M_x(\partial H_z/\partial x) + M_y(\partial H_z/\partial y) + M_z(\partial H_z/\partial z) - (\partial M_x/\partial t)\varepsilon_o E_y + (\partial M_y/\partial t)\varepsilon_o E_x. \tag{C1}$$

Substituting from Eqs.(A1) into Eq.(C1), integrating the force over the cross-sectional area of the beam, using the identity in Eq.(A7), and finally separating the terms in which $\omega' = -\omega$ from those in which $\omega' \neq -\omega$, we obtain

$$\iint F_z(x,y,z,t)\,dxdy = \tfrac{1}{2}\mathrm{Imag}\sum_{\omega_{1,2}}\iint \{\varepsilon_o(\varepsilon-1)(k_x\mathcal{E}_x + k_y\mathcal{E}_y + k_z\mathcal{E}_z)\mathcal{E}_z^* + \mu_o(\mu-1)(k_x\mathcal{H}_x + k_y\mathcal{H}_y + k_z\mathcal{H}_z)\mathcal{H}_z^*$$
$$+ [k_z\varepsilon_o(\varepsilon-1)/\mu](|\mathcal{E}_x|^2 + |\mathcal{E}_y|^2 + |\mathcal{E}_z|^2) + [k_z\mu_o(\mu-1)/\varepsilon](|\mathcal{H}_x|^2 + |\mathcal{H}_y|^2 + |\mathcal{H}_z|^2)\}\,dk_x dk_y$$
$$+ \tfrac{1}{4}\mathrm{i}\sum_{\omega'\neq-\omega}\iint\{\varepsilon_o(\varepsilon-1)[-k_x\mathcal{E}_x(k_x,k_y,\omega) - k_y\mathcal{E}_y(k_x,k_y,\omega) + k_z'\mathcal{E}_z(k_x,k_y,\omega)]\mathcal{E}_z(-k_x,-k_y,\omega')$$
$$+ \mu_o(\mu-1)[-k_x\mathcal{H}_x(k_x,k_y,\omega) - k_y\mathcal{H}_y(k_x,k_y,\omega) + k_z'\mathcal{H}_z(k_x,k_y,\omega)]\mathcal{H}_z(-k_x,-k_y,\omega')$$

$$-\omega\mu_o\varepsilon_o(\varepsilon-1)[\mathcal{E}_x(k_x,k_y,\omega)\mathcal{H}_y(-k_x,-k_y,\omega')-\mathcal{E}_y(k_x,k_y,\omega)\mathcal{H}_x(-k_x,-k_y,\omega')]$$

$$-\omega\mu_o\varepsilon_o(\mu-1)[\mathcal{E}_x(-k_x,-k_y,\omega')\mathcal{H}_y(k_x,k_y,\omega)-\mathcal{E}_y(-k_x,-k_y,\omega')\mathcal{H}_x(k_x,k_y,\omega)]\}$$

$$\times\exp[i(k_z+k_z')z]\exp[-i(\omega+\omega')t]dk_x dk_y. \qquad (C2)$$

In the above equation, the first sum (corresponding to $\omega'=-\omega$) vanishes because, according to Maxwell's 1st and 4th equations both $\mathbf{k}\cdot\mathcal{E}$ and $\mathbf{k}\cdot\mathcal{H}$ are zero; moreover, the remaining terms are all real-valued. As for the second sum, we integrate this force along the $z$-axis from $z=0$ to $V_g t$, i.e., over the length of the beat waveform that enters the medium during the time interval $(0, t)$. Here $V_g=(\omega_1-\omega_2)/(k_{z1}-k_{z2})$ is the group velocity, which, in the limit of small $k_x$ and $k_y$, approaches the expression in Eq.(A5). Next we integrate over a single beat period, from $t=0$ to $T=2\pi/(\omega_2-\omega_1)$, noting that $(1/T)\int_0^T \exp[-i(\omega+\omega')t]dt = 0$ for all allowed combinations of $\omega$ and $\omega'$ (since $\omega'=-\omega$ is already excluded from the second sum). The second sum in Eq.(C2) contains 12 different combinations of $\pm\omega_1$ and $\pm\omega_2$, but only four $(\omega,\omega')$ pairs which have opposite signs, namely, $(-\omega_1,\omega_2)$, $(-\omega_2,\omega_1)$, $(\omega_1,-\omega_2)$, $(\omega_2,-\omega_1)$, contribute to $<F_z>$. For these pairs $(k_z+k_z')V_g\approx(\omega+\omega')$ and, therefore,

$$(1/T)\int_0^T \exp\{i[(k_z+k_z')V_g-(\omega+\omega')]t\}dt = 1. \qquad (C3)$$

As for the remaining eight pairs such as $(\omega_1,\omega_1)$ or $(-\omega_1,-\omega_2)$ which have identical signs, we note that $(k_z+k_z')V_g-(\omega+\omega')\approx 2\omega[(n_p/n_g)-1]$. Now, if the phase index $n_p$ happens to differ substantially from the group index $n_g$, the corresponding exponential function will vary rapidly with time, rendering the time-averaged contributions of the remaining eight terms negligible. If, on the other hand, $n_p$ and $n_g$ happen to be so close as to cause the exponential function to vary only slowly during the beat period, the eight terms split into two groups of four positive and four negative terms. [According to Eqs.(A4), the terms arising from either $(\omega_1,\omega_2)$ or $(\omega_2,\omega_1)$ are equal in magnitude but opposite in sign compared to those arising from $(\omega_1,\omega_1)$ and $(\omega_2,\omega_2)$]. The two groups of terms thus cancel each other out. Therefore, under all circumstances, only four terms contribute to the time-averaged force $<F_z>$, yielding

$$<F_z> = (1/AT)\int_0^T\int_0^{V_g t}\iint_{-\infty}^{\infty} F_z(x,y,z,t)dxdydzdt$$

$$= (1/4A)\sum_{\pm\omega_{1,2}}\iint\{\varepsilon_o(\varepsilon-1)\mathcal{E}_z(k_x,k_y,\omega)\mathcal{E}_z^*(k_x,k_y,-\omega')+\mu_o(\mu-1)\mathcal{H}_z(k_x,k_y,\omega)\mathcal{H}_z^*(k_x,k_y,-\omega')$$

$$-\omega\mu_o\varepsilon_o(\varepsilon-1)(k_z+k_z')^{-1}[\mathcal{E}_x(k_x,k_y,\omega)\mathcal{H}_y^*(k_x,k_y,-\omega')-\mathcal{E}_y(k_x,k_y,\omega)\mathcal{H}_x^*(k_x,k_y,-\omega')]$$

$$-\omega\mu_o\varepsilon_o(\mu-1)(k_z+k_z')^{-1}[\mathcal{E}_x^*(k_x,k_y,-\omega')\mathcal{H}_y(k_x,k_y,\omega)-\mathcal{E}_y^*(k_x,k_y,-\omega')\mathcal{H}_x(k_x,k_y,\omega)]\}dk_x dk_y.$$

$$(C4)$$

Since, according to Eqs.(A4), the field amplitudes of the beams with frequencies $\omega_1$ and $\omega_2$ are equal but opposite in sign, Eq.(C4) reduces to

$$<F_z> = -(2A)^{-1}\text{Real}\sum_{\omega_{1,2}}\iint\left[\varepsilon_o(\varepsilon-1)|\mathcal{E}_z|^2+\mu_o(\mu-1)|\mathcal{H}_z|^2-\frac{\omega\mu_o\varepsilon_o(\varepsilon+\mu-2)}{k_z+k_z'}(\mathcal{E}_x\mathcal{H}_y^*-\mathcal{E}_y\mathcal{H}_x^*)\right]dk_x dk_y.$$

$$(C5)$$

In the limit when the beam's cross-sectional area $A\to\infty$, the integrals of $|\mathcal{E}_z|^2$ and $|\mathcal{H}_z|^2$ approach zero, while $k_z+k_z'\to(\omega_1/c)\sqrt{\mu_1\varepsilon_1}-(\omega_2/c)\sqrt{\mu_2\varepsilon_2}$ for $(\omega,\omega')=(\omega_1,-\omega_2)$; the sign of the expression is reversed for $(\omega,\omega')=(\omega_2,-\omega_1)$. Replacing for $\mathcal{E}_x,\mathcal{E}_y,\mathcal{H}_x,\mathcal{H}_y$ in terms of the incident field amplitudes $\mathcal{E}_{xo},\mathcal{E}_{yo},\mathcal{H}_{xo},\mathcal{H}_{yo}$ and the Fresnel reflection coefficients $r_1, r_2$ (given by Eq.(11)), we find the time-averaged force per unit surface area of the slab to be

$$\langle F_z \rangle = (1+r^2)\text{Real}\left\{(Ac)^{-1}\iint_{-\infty}^{\infty}(\mathcal{E}_{xo}\mathcal{H}_{yo}^* - \mathcal{E}_{yo}\mathcal{H}_{xo}^*)\,\mathrm{d}k_x\mathrm{d}k_y\right\}$$

$$-V_g\,\text{Real}\left\{(Ac^2)^{-1}\iint_{-\infty}^{\infty}(\mathcal{E}_x\mathcal{H}_y^* - \mathcal{E}_y\mathcal{H}_x^*)\,\mathrm{d}k_x\mathrm{d}k_y\right\}. \tag{C6}$$

In this equation, the first term is the time-rate of flow of incident plus reflected momenta in the free space, while the second term accounts for the influx of Abraham momentum into the transparent medium. We have thus proved the conservation of linear momentum upon reflection from a transparent, dispersive, magnetic dielectric.

**Appendix D: Torque exerted on a dispersive slab**

We compute the torque experienced by the slab of Fig. 1 when the beat-waveform of Eqs. (A1) enters the medium. Only an outline of the derivation will be given here as the methods employed are similar to those that have already been described. Starting with Eq. (5), we write

$$T_z(x,y,z,t) = x[P_x(\partial E_y/\partial x) + P_y(\partial E_y/\partial y) + P_z(\partial E_y/\partial z) + (\partial P_z/\partial t)\mu_o H_x - (\partial P_x/\partial t)\mu_o H_z$$

$$+ M_x(\partial H_y/\partial x) + M_y(\partial H_y/\partial y) + M_z(\partial H_y/\partial z) - (\partial M_z/\partial t)\varepsilon_o E_x + (\partial M_x/\partial t)\varepsilon_o E_z]$$

$$- y[P_x(\partial E_x/\partial x) + P_y(\partial E_x/\partial y) + P_z(\partial E_x/\partial z) + (\partial P_y/\partial t)\mu_o H_z - (\partial P_z/\partial t)\mu_o H_y + M_x(\partial H_x/\partial x)$$

$$+ M_y(\partial H_x/\partial y) + M_z(\partial H_x/\partial z) - (\partial M_y/\partial t)\varepsilon_o E_z + (\partial M_z/\partial t)\varepsilon_o E_y]$$

$$+ (P_x E_y - P_y E_x) + (M_x H_y - M_y H_x). \tag{D1}$$

Expanding each field in its Fourier integral, combining the various terms, then separating the terms in which $\omega' = -\omega$ from those in which $\omega' \neq -\omega$, we obtain

$$\iint T_z(x,y,z,t)\,\mathrm{d}x\mathrm{d}y = \tfrac{1}{4}\sum_{\omega_{1,2}}\iint\left\{(k_y\partial/\partial k_x - k_x\partial/\partial k_y)[\varepsilon_o(\varepsilon-1)(|\mathcal{E}_x|^2+|\mathcal{E}_y|^2+|\mathcal{E}_z|^2)\right.$$

$$\left. + \mu_o(\mu-1)(|\mathcal{H}_x|^2+|\mathcal{H}_y|^2+|\mathcal{H}_z|^2)]\right\}\mathrm{d}k_x\mathrm{d}k_y$$

$$+ \tfrac{1}{4}\sum_{\omega'\neq-\omega}\iint\left\{\varepsilon_o(\varepsilon'-1)[k_y\mathcal{E}_x\partial\mathcal{E}_x'^*/\partial k_x - k_x\mathcal{E}_x\partial\mathcal{E}_x'^*/\partial k_y + k_y\mathcal{E}_y\partial\mathcal{E}_y'^*/\partial k_x - k_x\mathcal{E}_y\partial\mathcal{E}_y'^*/\partial k_y - (k_zk_z')^{-1}\right.$$

$$\times (k_x\mathcal{E}_x + k_y\mathcal{E}_y)(k_y\mathcal{E}_x'^* - k_x\mathcal{E}_y'^* + k_xk_y\partial\mathcal{E}_x'^*/\partial k_x + k_y^2\partial\mathcal{E}_y'^*/\partial k_x - k_x^2\partial\mathcal{E}_x'^*/\partial k_y - k_xk_y\partial\mathcal{E}_y'^*/\partial k_y)]$$

$$+ \mu_o(\mu'-1)[k_y\mathcal{H}_x\partial\mathcal{H}_x'^*/\partial k_x - k_x\mathcal{H}_x\partial\mathcal{H}_x'^*/\partial k_y + k_y\mathcal{H}_y\partial\mathcal{H}_y'^*/\partial k_x - k_x\mathcal{H}_y\partial\mathcal{H}_y'^*/\partial k_y - (k_zk_z')^{-1}$$

$$\times (k_x\mathcal{H}_x + k_y\mathcal{H}_y)(k_y\mathcal{H}_x'^* - k_x\mathcal{H}_y'^* + k_xk_y\partial\mathcal{H}_x'^*/\partial k_x + k_y^2\partial\mathcal{H}_y'^*/\partial k_x - k_x^2\partial\mathcal{H}_x'^*/\partial k_y - k_xk_y\partial\mathcal{H}_y'^*/\partial k_y)]$$

$$+ \varepsilon_o(\varepsilon-1)(\mathcal{E}_x\mathcal{E}_y'^* - \mathcal{E}_y\mathcal{E}_x'^*) - \varepsilon_o(\mu-1)(\mu_o\varepsilon_o\mu\mu'\omega\omega'k_zk_z')^{-1}[k_xk_y(k_z'^2-k_z^2)(\mathcal{E}_x\mathcal{E}_x'^* - \mathcal{E}_y\mathcal{E}_y'^*)$$

$$- (k_z^2k_z'^2 + k_x^2k_z'^2 + k_y^2k_z^2)\mathcal{E}_x\mathcal{E}_y'^* + (k_z^2k_z'^2 + k_x^2k_z^2 + k_y^2k_z'^2)\mathcal{E}_y\mathcal{E}_x'^*]$$

$$+ \varepsilon_o(\mu\varepsilon'-1)(\omega+\omega')(\mu\omega)^{-1}\left\{(k_x\mathcal{E}_y - k_y\mathcal{E}_x)(\partial\mathcal{E}_x'^*/\partial k_x + \partial\mathcal{E}_y'^*/\partial k_y) + (k_zk_z'^3)^{-1}\right.$$

$$\times [k_xk_y(k_z'^2-k_z^2)(\mathcal{E}_x\mathcal{E}_x'^* - \mathcal{E}_y\mathcal{E}_y'^*) - (k_z^2k_z'^2 + k_x^2k_z'^2 + k_y^2k_z^2)\mathcal{E}_x\mathcal{E}_y'^*$$

$$+ (k_z^2k_z'^2 + k_x^2k_z^2 + k_y^2k_z'^2)\mathcal{E}_y\mathcal{E}_x'^*] + (k_zk_z')^{-1}[k_xk_y\mathcal{E}_x + (k_y^2+k_z^2)\mathcal{E}_y](k_x\partial\mathcal{E}_x'^*/\partial k_x + k_y\partial\mathcal{E}_y'^*/\partial k_x)$$

$$- (k_zk_z')^{-1}[(k_x^2+k_z^2)\mathcal{E}_x + k_xk_y\mathcal{E}_y](k_x\partial\mathcal{E}_x'^*/\partial k_y + k_y\partial\mathcal{E}_y'^*/\partial k_y)\left.\right\}\right\}$$

$$\times \exp[i(k_z+k_z')z]\exp[-i(\omega+\omega')t]\,\mathrm{d}k_x\mathrm{d}k_y. \tag{D2}$$

In the above equation, $\mathcal{E}$ and $\mathcal{E}'$ stand for $\mathcal{E}(k_x,k_y,\omega)$ and $\mathcal{E}(k_x,k_y,-\omega')$, respectively, while $(\mu,\varepsilon)$ and $(\mu',\varepsilon')$ are the material parameters at $\omega$ and $\omega'$. The finite diameter of the $\mathcal{E}$- and $\mathcal{H}$-field profiles in the $k_xk_y$-plane makes the first sum in Eq. (D2) exactly equal to zero. We integrate the second sum over $z$ (from 0 to $V_gt$) and then again over $t$ (from 0 to $T$) to obtain:

$$\langle T_z\rangle = (1/AT)\int_0^T\!\!\int_0^{V_g t}\!\!\iint_{-\infty}^{\infty} T_z(x,y,z,t)\,\mathrm{d}x\mathrm{d}y\mathrm{d}z\mathrm{d}t = (2A)^{-1}\mathrm{Imag}\sum_{\omega_{1,2}}\!\!\iint_{-\infty}^{\infty}(k_z+k_z')^{-1}\{\varepsilon_o(\varepsilon'-1)$$

$$\times[k_y\mathcal{E}_x\partial\mathcal{E}_x'^*/\partial k_x - k_x\mathcal{E}_x\partial\mathcal{E}_x'^*/\partial k_y + k_y\mathcal{E}_y\partial\mathcal{E}_y'^*/\partial k_x - k_x\mathcal{E}_y\partial\mathcal{E}_y'^*/\partial k_y - (k_z k_z')^{-1}(k_x\mathcal{E}_x+k_y\mathcal{E}_y)$$

$$\times(k_y\mathcal{E}_x'^* - k_x\mathcal{E}_y'^* + k_x k_y \partial\mathcal{E}_x'^*/\partial k_x + k_y^2\partial\mathcal{E}_y'^*/\partial k_x - k_x^2\partial\mathcal{E}_x'^*/\partial k_y - k_x k_y\partial\mathcal{E}_y'^*/\partial k_y)]$$

$$+\mu_o(\mu'-1)[k_y\mathcal{H}_x\partial\mathcal{H}_x'^*/\partial k_x - k_x\mathcal{H}_x\partial\mathcal{H}_x'^*/\partial k_y + k_y\mathcal{H}_y\partial\mathcal{H}_y'^*/\partial k_x - k_x\mathcal{H}_y\partial\mathcal{H}_y'^*/\partial k_y - (k_z k_z')^{-1}$$

$$\times(k_x\mathcal{H}_x+k_y\mathcal{H}_y)(k_y\mathcal{H}_x'^* - k_x\mathcal{H}_y'^* + k_x k_y\partial\mathcal{H}_x'^*/\partial k_x + k_y^2\partial\mathcal{H}_y'^*/\partial k_x - k_x^2\partial\mathcal{H}_x'^*/\partial k_y - k_x k_y\partial\mathcal{H}_y'^*/\partial k_y)]$$

$$+\varepsilon_o(\varepsilon-1)(\mathcal{E}_x\mathcal{E}_y'^* - \mathcal{E}_y\mathcal{E}_x'^*) - \varepsilon_o(\mu-1)(\mu_o\varepsilon_o\mu\mu'\omega\omega' k_z k_z')^{-1}[k_x k_y(k_z'^2-k_z^2)(\mathcal{E}_x\mathcal{E}_x'^* - \mathcal{E}_y\mathcal{E}_y'^*)$$

$$-(k_z^2 k_z'^2 + k_x^2 k_z'^2 + k_y^2 k_z^2)\mathcal{E}_x\mathcal{E}_y'^* + (k_z^2 k_z'^2 + k_x^2 k_z^2 + k_y^2 k_z'^2)\mathcal{E}_y\mathcal{E}_x'^*]$$

$$+\varepsilon_o(\mu\varepsilon'-1)(\omega+\omega')(\mu\omega)^{-1}\{(k_x\mathcal{E}_y - k_y\mathcal{E}_x)(\partial\mathcal{E}_x'^*/\partial k_x + \partial\mathcal{E}_y'^*/\partial k_y) + (k_z k_z'^3)^{-1}[k_x k_y(k_z'^2-k_z^2)$$

$$\times(\mathcal{E}_x\mathcal{E}_x'^* - \mathcal{E}_y\mathcal{E}_y'^*) - (k_z^2 k_z'^2 + k_x^2 k_z'^2 + k_y^2 k_z^2)\mathcal{E}_x\mathcal{E}_y'^* + (k_z^2 k_z'^2 + k_x^2 k_z^2 + k_y^2 k_z'^2)\mathcal{E}_y\mathcal{E}_x'^*]$$

$$+(k_z k_z')^{-1}[k_x k_y\mathcal{E}_x + (k_y^2 + k_z^2)\mathcal{E}_y](k_x\partial\mathcal{E}_x'^*/\partial k_x + k_y\partial\mathcal{E}_y'^*/\partial k_x)$$

$$-(k_z k_z')^{-1}[(k_x^2 + k_z^2)\mathcal{E}_x + k_x k_y\mathcal{E}_y](k_x\partial\mathcal{E}_x'^*/\partial k_y + k_y\partial\mathcal{E}_y'^*/\partial k_y)\}\}\,\mathrm{d}k_x\mathrm{d}k_y. \qquad (D3)$$

In the absence of any *orbital* angular momentum, the symmetry of the beam in the cross-sectional $xy$-plane ensures that the derivatives in Eq. (D3) drop out. We are then left with

$$\langle T_z\rangle \approx (\varepsilon_o/2A)\mathrm{Imag}\sum_{\omega_{1,2}}\!\!\iint(k_z+k_z')^{-1}\{(\varepsilon-1)(\mathcal{E}_x\mathcal{E}_y'^* - \mathcal{E}_y\mathcal{E}_x'^*)$$

$$+(\mu\omega k_z k_z'^3)^{-1}[(\mu\varepsilon'-1)(\omega+\omega') - k_z'^2(\mu-1)(\mu_o\varepsilon_o\mu'\omega')^{-1}][k_x k_y(k_z'^2-k_z^2)(\mathcal{E}_x\mathcal{E}_x'^* - \mathcal{E}_y\mathcal{E}_y'^*)$$

$$-(k_z^2 k_z'^2 + k_x^2 k_z'^2 + k_y^2 k_z^2)\mathcal{E}_x\mathcal{E}_y'^* + (k_z^2 k_z'^2 + k_x^2 k_z^2 + k_y^2 k_z'^2)\mathcal{E}_y\mathcal{E}_x'^*]\}\,\mathrm{d}k_x\mathrm{d}k_y. \qquad (D4)$$

Here the summation contains only two $(\omega,\omega')$ pairs, namely, $(\omega_1,-\omega_2)$ and $(\omega_2,-\omega_1)$. In the limit when the beam has a large cross-sectional area $A$, terms in Eq. (D4) that contain $k_x$ and/or $k_y$ make negligible contributions to the integral and can safely be ignored. We set $k_z\approx(\omega/c)\sqrt{\mu\varepsilon}$, $k_z'\approx(\omega'/c)\sqrt{\mu'\varepsilon'}$, and assume a circularly-polarized beam having $\mathcal{E}_y=\mathrm{i}\mathcal{E}_x$. Accounting for the fact that $\mathcal{E}_x$ and $\mathcal{E}_x'$ have equal magnitudes and opposite signs, in the limit when $\omega_1\to\omega_2$, Eq. (D4) may be written

$$\langle T_z\rangle \approx 2c(\varepsilon_o/A)\frac{\{\mu\varepsilon + \tfrac{1}{2}\omega[\varepsilon(\mathrm{d}\mu/\mathrm{d}\omega)+\mu(\mathrm{d}\varepsilon/\mathrm{d}\omega)]\}-1}{\omega\mu[\mathrm{d}(\omega\sqrt{\mu\varepsilon})/\mathrm{d}\omega]}\iint_{-\infty}^{\infty}|\mathcal{E}_x(k_x,k_y,\omega)|^2\,\mathrm{d}k_x\mathrm{d}k_y. \qquad (D5)$$

Here $\mathrm{d}\varepsilon/\mathrm{d}\omega = [\varepsilon(\omega_1)-\varepsilon(\omega_2)]/(\omega_1-\omega_2)$ and $\mathrm{d}\mu/\mathrm{d}\omega = [\mu(\omega_1)-\mu(\omega_2)]/(\omega_1-\omega_2)$. The expression for time-averaged torque per unit area can finally be written in compact form as follows:

$$\langle T_z\rangle \approx 2(\varepsilon_o/A\omega)(c\sqrt{\varepsilon/\mu} - \mu^{-1}V_g)\iint_{-\infty}^{\infty}|\mathcal{E}_x(k_x,k_y,\omega)|^2\,\mathrm{d}k_x\mathrm{d}k_y. \qquad (D6)$$

In the above equation, the first term represents $(1-r^2)$ times the rate of flow of incident angular momentum in the free space, while the second term accounts for the electromagnetic (i.e., Abraham) angular momentum that pours into the medium with the group velocity $V_g$. The equation therefore confirms the conservation of angular momentum.

**Acknowledgement**

This work has been supported by the Air Force Office of Scientific Research (AFOSR) under contract number FA 9550–04–1–0213.